\documentclass[10pt,prl,aps,twocolumn,showpacs,superscriptaddress,floatfix]{revtex4}
\usepackage{graphicx}
\usepackage{bm}
\usepackage{amssymb}
\usepackage{color}

\begin{document}

\title{
Topological Insulating States in Laterally Patterned Ordinary  Semiconductors
}

\author{O. P. Sushkov}
\affiliation{School of Physics, University of New South Wales, Sydney 2052, 
Australia}
\author{A. H. Castro Neto}
\affiliation{Graphene Research Centre and Physics Department, National University of Singapore, 6 Science Drive 2, Singapore 117546}

\begin{abstract}
We propose that ordinary semiconductors with large spin-orbit coupling (SOC), such as GaAs, can host stable, robust, and {\it tunable} 
topological states in the presence of quantum confinement and superimposed potentials with hexagonal symmetry. 
 We show that
the electronic gaps which support chiral spin edge states 
can be as large as the electronic bandwidth 
in the heterostructure miniband.
The existing lithographic technology can produce  a topological insulator (TI) operating 
at temperature  $10- 100K$. Improvement of lithographic techniques  will open way to
tunable room temperature TI.

\end{abstract}



\date{\today}

\maketitle

A topological insulator (TI) is a fascinating state of matter which presents unusual physical properties such as a quantum Spin Hall effect 
in two-dimensions (2D), spin-polarized chiral Dirac surface states in three-dimensions (3D), exotic magneto-electric effects, and Majorana 
fermions in the presence of superconductivity \cite{zhang,hasankane,moore}. Although 
TIs have  attracted a lot of attention, progress in this area has been hindered by the absence of 
experimental systems with robust electronic and structural properties. Although these materials have been christened as ``topologically 
protected'', material issues associated the weak strength of the SOC, the small size of the gaps, and the strong disorder (of the order of 
the electronic bandwidth) present in most of the proposed systems, make the experimental realization of these amazing physical properties 
very difficult, if not unattainable. 

One of the first theoretical predictions for such TI states was made for graphene \cite{Kane05} in the early days of graphene research 
\cite{Greview}.
Nevertheless, carbon is a light element with weak intrinsic SOC ($\approx 10^{-2}$ eV). In addition, in flat graphene, 
due to wavefunction orthogonality, the SOC for the $\pi$-bands is even weaker ($\approx 10^{-6}$ eV) and, 
therefore, essentially unobservable. While deviations from flat sp$^2$, to out of plane sp$^3$, bonds can lead to three fold 
enhancement of the SOC \cite{netoguinea}, the atomic control over these deformations is a major experimental challenge. 
Shortly after its initial proposal, TI states where predicted to occur in HgTe quantum wells \cite{hgte}, in 3D bulk solids of binary compounds 
involving Bi \cite{biti}, and half-Heusler ternary compounds \cite{hsinlin}. 
Unfortunately, all these materials 
are very sensitive to stoichiometry.
 Hence, the unavoidable presence of defects, which 
are usually unitary scatterers and can act as donors/acceptors, has a strong effect in the electronic structure. 
One observes, for instance, large broadening of the spectral lines for surface states 
in angle resolved  photo-emission, and doping of the bulk crystal ultimately transforming the TIs into metals \cite{kanemoore}. 

 An idea to use semiconductors to produce TI was put forward in Ref. \cite{Liu}.
In particular it was suggested to use inverted InAs/GaSb Quantum Wells.
The system was realized experimentally with some indications for  helical edge modes \cite{knez}.
In this paper, we propose an alternative way to produce robust, structurally stable, and {\it tunable} TI states in ordinary semiconductors
such as GaAs. The advantages of these materials are that they have significant SOC, they can be grown with extreme precision using
molecular beam epitaxy (MBE) with large electronic mobilities, they can be tailored into quantum wells with arbitrary 
thickness, and can be controlled by external gates \cite{socreview}. The robustness and flexibility of these systems can be
the starting point for the creation of different kinds of TIs that cannot be obtained otherwise. 

In fact, it is known that hole-doped zinc-blend semiconductors naturally have large SOC that originates from the atomic $p_{3/2} - p_{1/2}$ 
fine structure splitting. 
We demonstrate that the effective SOC in a semiconductor quantum well with a superimposed hexagonal superlattice can be 
controlled by the strength of the transverse confinement and the scale of the superlattice. Hence, the SOC gap can made comparable 
to the bandwidth or continuously switched to zero. 
Finally, we show that the SOC leads to the appearance of chiral spin edge modes in contrast with systems such as graphene where these
edge states exist even in the absence of SOC \cite{Gedge}. Thus, in our proposal, the system can be {\it continuously tuned} between the 
Dirac metal, topological insulator, and  standard band insulator.

As it is well known \cite{socreview}, in the 3D bulk of a semiconductor, 
the hole wave function in systems like GaAs originates from the atomic $p_{3/2}$ 
orbital and  thus, the hole has an angular momentum $J=3/2$ (the so-called, hole spin $S=3/2$). In the large wavelength approximation 
(the ${\bf k} \cdot {\bf p}$ approximation), the hole effective Hamiltonian is proportional to the second power of the hole momentum
${\bm k}$. The only kinematic structures allowed by symmetries are ${\bm k}^2$, $({\bm k}\cdot{\bm S})^2$, and
$T_{\mu\nu\alpha\gamma}k_{\mu}k_{\nu}S_{\alpha}S_{\gamma}$, where $T_{\mu\nu\alpha\gamma}$ is the 4$^{\rm th}$ rank tensor built of unit vectors
of the cubic lattice that is known to be parametrically small relative to the other terms \cite{socreview} and will be disregarded 
in what follows. In this case, the effective Hamiltonian can be written as (we use units such that $\hbar =1$): 
\begin{equation}
\label{H}
H_{3D}=\frac{k^2}{2 m_e}\left(\gamma_1+2.5\tilde\gamma\right)-
\tilde\gamma\frac{({\bm k}\cdot{\bm S})^2}{m_e} \ ,
\end{equation}
where $m_e$ is the free electron mass and $\gamma_1$, $\tilde\gamma$ are the Luttinger-Kohn parameters~\cite{Luttinger}. 
Hereafter, we use the parameters for GaAs where $\gamma_1\approx 6.8$, $\tilde\gamma \approx 2.9$~\cite{gval}. 
Notice that $\tilde\gamma$ parametrizes the effective SOC and is comparable with $\gamma_1$, which parametrizes 
the effective hole kinetic energy. 

The 3D semiconductor can be geometrically confined in one direction creating at 2D quantum well. For simplicity,
we assume that the 
confinement is described by an infinite square well of width $d$ and that only the lowest quantum state, $|0\rangle$, has to be 
taken into account. In this case we have 
$\langle 0|{\hat k}_z|0\rangle =0$, and $\langle 0|{\hat k}_z^2|0 \rangle =\pi^2/d^2$.
As a result, the in-plane momentum is small $k_x,k_y \ll \pi/d$ and the $k_z^2S_z^2$ term in (\ref{H}) enforces the spin quantization 
along the z-axis. The lowest energy state corresponds to $S_z=\pm 3/2$ and the higher state corresponds to $S_z=\pm 1/2$ giving
rise to the heavy ($S_z=\pm 3/2$) and light ($S_z=\pm 1/2$) hole states \cite{socreview}. According to (\ref{H}) the value of the 
splitting between these states is given by:
$\Delta=E_{1/2}-E_{3/2}=4\tilde\gamma\frac{(\pi/d)^2}{2m_e}$ . 
When the hole density is low only the heavy hole band is filled. The heavy hole 2D dispersion follows from (\ref{H}) and
has several contributions. Firstly, there is a diagonal term due to the z-confinement contribution
which is just the matrix element of (\ref{H}):
\begin{eqnarray}
\label{H2d}
\frac{p^2}{2 m_e}\left(\gamma_1+2.5\tilde\gamma\right)-
\tilde\gamma\frac{p^2}{2 m_e}\langle3/2|S_x^2|3/2\rangle
=\frac{p^2}{2 m_e}\left(\gamma_1+\tilde\gamma\right)\ 
\end{eqnarray}
Here ${\bm p}=(k_x,k_y)$ is the in-plane momentum.
There is also the 2$^{\rm nd}$ order perturbation theory contribution due to
the $-\frac{\tilde\gamma}{m_e}[k_xk_z(S_xS_z+S_zS_x)+k_yk_z(S_yS_z+S_zS_y)]$ term
in (\ref{H}). This term generates virtual z-excitations.
A straightforward calculation gives the following 2$^{\rm nd}$ order contribution for GaAs: $-1.6\frac{p^2}{2 m_e}$.
Putting these results together with (\ref{H2d}) we find that the in-plane mass of the heavy hole is
$m^*=m_e/\left(\gamma_1+\tilde\gamma-1.6\right)\approx 0.12 m_e$ \ .
For a soft parabolic z-confinement the mass can be somewhat larger, $m^* \sim 0.15-0.2 m_e$.

The $({\bm p}\cdot{\bm S})^2$ part of the Hamiltonian (\ref{H}) leads to the heavy-light hole mixing.
The mixing matrix elements are:
\begin{eqnarray}
\label{me}
&&\langle {\bm p},-1/2|H|{\bm p},3/2\rangle=
-\frac{\sqrt{3}\tilde\gamma}{2m_e}(p_x+ip_y)^2\nonumber\\
&&\langle {\bm p},1/2|H|{\bm p},-3/2\rangle=
-\frac{\sqrt{3}\tilde\gamma}{2m_e}(p_x-ip_y)^2\nonumber \, ,
\end{eqnarray}
with states given by:
\begin{eqnarray}
\label{Se}
&&|{\bm p},\uparrow\rangle=\left[|+\frac{3}{2}\rangle
+\alpha (p_x+ip_y)^2|-\frac{1}{2}\rangle \right]e^{i{\bm p}\cdot{\bm r}}\nonumber\\
&&|{\bm p},\downarrow\rangle=\left[|-\frac{3}{2}\rangle
+\alpha (p_x-ip_y)^2|+\frac{1}{2}\rangle \right]e^{i{\bm p}\cdot{\bm r}}\nonumber\\
&&\alpha=\frac{\sqrt{3}\tilde\gamma }{2m_e\Delta}=
\frac{\sqrt{3}d^2}{4\pi^2}\ ,\ \ \ \alpha^2p^4 \ll 1 \ .
\end{eqnarray}
Here we have introduced an effective ``spin'' $s=1/2$ degree of freedom describing the two $|\uparrow\rangle$ and $|\downarrow\rangle$ states.
Interestingly, $\tilde\gamma$, which can be considered as a strength of the spin orbit
interaction, see Eq. (\ref{H}), is cancelled out in the expression for $\alpha$ in Eq. (\ref{Se}).
One can call it  "ultrarelativistic" behaviour, the spin-orbit is so large that it does not
appear explicitly in the answer.

\begin{figure}[ht]
\includegraphics[width=0.2\textwidth,clip]{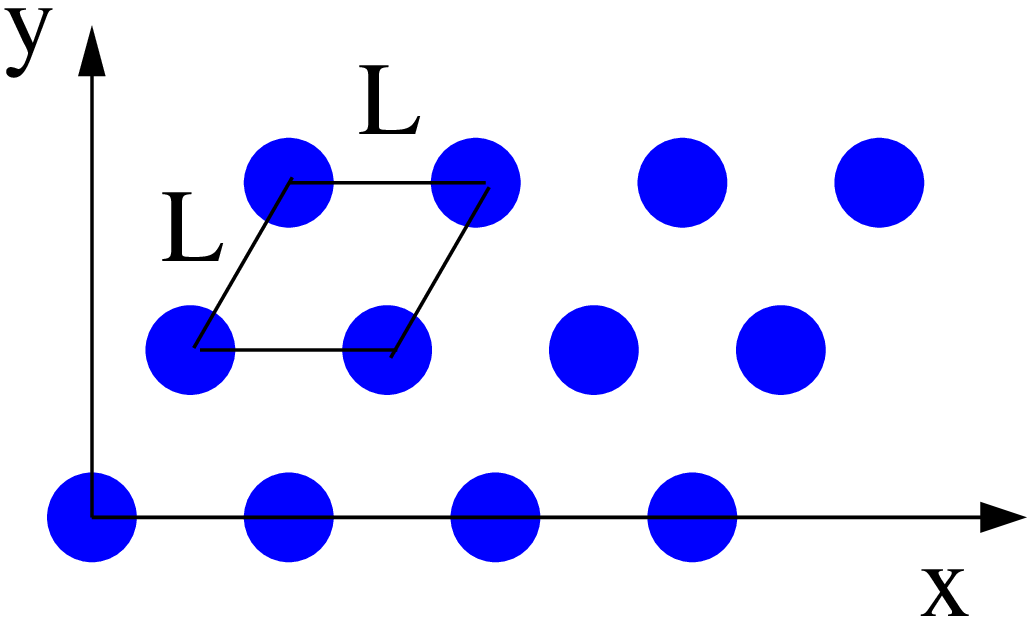}
\hspace{20pt}
\includegraphics[width=0.15\textwidth,clip]{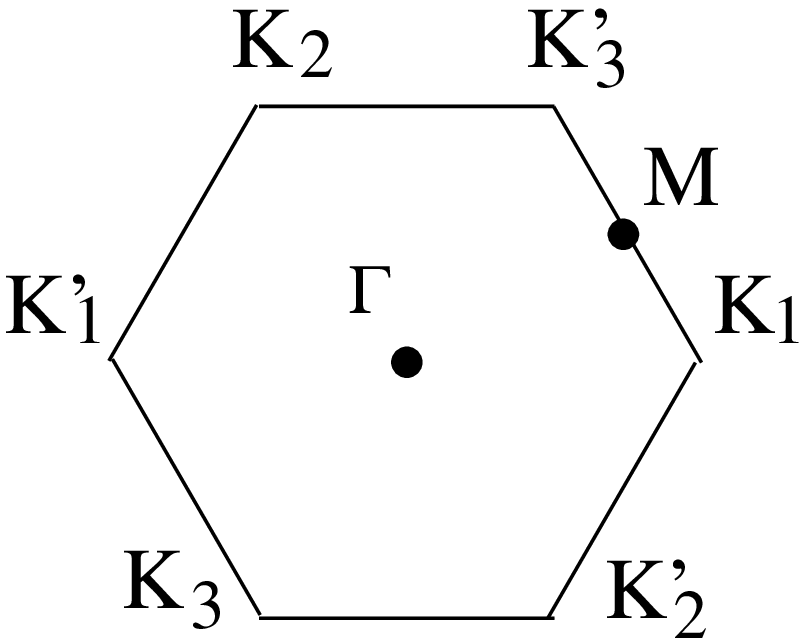}
\caption{ Triangular lattice (left) and the corresponding Brillouin 
zone (right).
}
\label{tri}
\end{figure}

In order to generate the TI, a potential $U({\bm r})$ with hexagonal (triangular) symmetry and spacing $L$, as shown in Fig.\ref{tri},
is superimposed to the 2D electron gas \cite{Park09,Gibertini09,Singha11}.
The lattice translation vectors are ${\bm L}_1=(L,0)$ , ${\bm L}_2=(\frac{L}{2},\frac{\sqrt{3}L}{2})$.
Hence, there are two independent reciprocal lattice vectors in the Brillouin zone (see Fig.\ref{tri}):
\begin{eqnarray}
\label{trvi}
{\bm G}_1=\frac{2\pi}{3L}(3,\sqrt{3}), \ \ 
{\bm G}_2=\frac{2\pi}{3L}(0,2\sqrt{3}), \ \
{\bm G}_3={\bm G}_1-{\bm G}_2,\nonumber \, .
\end{eqnarray}
The points ${\bm K}_1$, ${\bm K}_2$, ${\bm K}_3$ are connected by vectors ${\bm G}_i$, and ${\bm K}_i'$ 
are obtained from the ${\bm K}_i$ by reflection. In order to simplify notation, we will measure energy
in units of the bandwidth:
\begin{eqnarray}
\label{e0}
E_0=\frac{K^2}{2m^*}=\frac{(4\pi/3L)^2}{2m^*} \ .
\end{eqnarray}
In the case of GaAs, assuming $L = 20$ nm, which can be obtained experimentally with standard 
lithographic techniques, we have $E_0 \approx 13$ meV.
Notice, however, this energy scale can be easily controlled by tuning $L$ (for $L =50$ nm, $E_0 \approx 2$ meV.

\begin{figure}[ht]
\includegraphics[width=0.18\textwidth,clip]{disp1.eps}
\includegraphics[width=0.2\textwidth,clip]{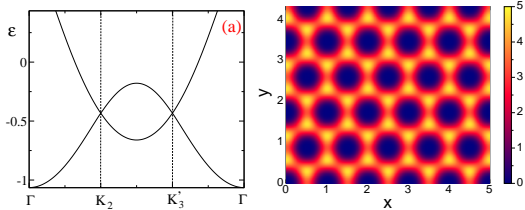}
\caption{(a) The hole dispersion along a particular contour in the
BZ, (b) Map of the total charge density (in units $1/L^2$) at the chemical potential
tuned to the Dirac point. Both figures correspond to ${\overline W}=W/E_0=1$ in the absence of SOC.}
\label{disp}
\end{figure}

Unlike the case of graphene where the starting point is a tight-binding description \cite{Greview}, our description starts
from a nearly free electron description. We assume, for simplicity, a periodic potential with a single Fourier component: 
\begin{equation}
\label{uh}
U({\bm r})=2W\left[\cos({\bm G}_1\cdot {\bm r})+\cos({\bm G}_2\cdot {\bm r})
+\cos({\bm G}_3\cdot {\bm r})\right] \, ,
\end{equation}
where $W$ gives the strength of the potential. This potential has nonzero matrix elements only between states 
$|{\bm k}\rangle$ and $|{\bm k}\pm {\bm G}_i \rangle$ with matrix elements given by $W$. 
Diagonalization of the Hamiltonian,  
\begin{equation}
\label{H2}
H=\frac{p^2}{2m^*}+U({\bm r}) \, ,
\end{equation}
gives the hole dispersion shown in Fig.~\ref{disp}(a) with the presence of two Dirac points with linear dispersion.   
If the chemical potential is tuned at the Dirac point the average hole density is $\langle n \rangle =8/(3L^2)$. The map of the
charge density is shown in Fig.~\ref{disp}(b). At $L=50nm$ the average density is $1.1\times 10^{11}cm^{-2}$.
Even when the potential is strong, namely ${\overline W}=W/E_0=1-2$, the dispersion, 
Fig.~\ref{disp}(a), is rather close to the result obtained by perturbation theory. 
The charge density plot, Fig.~\ref{disp}(b), is fully connected with empty spots at positions of
the potential maximums. So, in clear contrast to graphene, at ${\overline W} \lesssim 2$ the system is much closer to the
nearly free electron regime than to the tight-binding one. 

Perturbative analysis of the system in the nearly free electron regime, at small ${\overline W}=W/E_0$, is straightforward~\cite{Park09}.
A hole state close to the Dirac point, $q \ll 1$, is described by degenerate perturbation theory as:
\begin{eqnarray}
\label{psi}
\psi_{\bm q}&\propto&
c_1 |1\rangle +c_2 |2\rangle +c_3 |3\rangle \nonumber \\
|j\rangle&=&e^{i({\bm K}_j+{\bm q})\cdot{\bm r}}\ .
\end{eqnarray}
In the basis of states $|j\rangle$  the Hamiltonian (\ref{H2}) is represented by $3\times 3$ matrices:
\begin{eqnarray}
\frac{p^2}{2m^*} &\to& \delta_{ik}\frac{({\bm K}_i+{\bm q})^2}{2m^*}\approx E_0+\delta_{ik}\frac{{\bm K}_i\cdot{\bm q}}{m^*}
\nonumber\\
U &\to& U_{ik}=W \ .
\nonumber
\end{eqnarray}
The eigen-energies of the U-matrix are $0,0,3W$. 
In order to project in the double degenerate subspace of $U$ we define \cite{Park09}:
\begin{eqnarray}
\label{cc}
|a\rangle =\left(
\begin{array}{c}
0\\
\frac{1}{\sqrt{2}}\\
-\frac{1}{\sqrt{2}}
\end{array}\right) \, , \ \ \ \ \ \
|b\rangle =\left(
\begin{array}{c}
\sqrt{\frac{2}{3}}\\
-\frac{1}{\sqrt{6}}\\
-\frac{1}{\sqrt{6}}
\end{array}\right) \, .
\end{eqnarray}
Projecting the kinetic energy to this basis and shifting the zero energy level to $E_0$ one finds:
\begin{eqnarray}
\label{mev}
&&\langle a|H|a\rangle\to\frac{({\bm K}_2+{\bm K}_3)\cdot{\bm q}}{2 m^*}=
-v q_x \, ,
\nonumber\\
&&\langle b|H|b\rangle=\frac{(4{\bm K}_1+{\bm K}_2+{\bm K}_3)\cdot{\bm q}}{6 m^*}=
+vq_x \, ,
\nonumber\\
&&\langle b|H|a\rangle=\frac{({\bm K}_3-{\bm K}_2)\cdot{\bm q}}{2\sqrt{3} m^*}=-v q_y \ ,
\end{eqnarray}
where 
$v= \frac{K}{2m^*} = \frac{2 \pi}{3 L m^*}$ ,
is the Fermi-Dirac velocity. Notice that the velocity is controlled by the lattice spacing. 
Hence, in the Pauli matrix (pseudo-spin) representation the effective low energy Hamiltonian reads:
\begin{eqnarray}
\label{HP1}
H&=&v\left(-\sigma_zq_x-\sigma_xq_y\right) \, .
\end{eqnarray}
One can perform the unitary transformation $H \to T^{\dag}HT$, where $T$ represents two subsequent $\pi/2$ rotations around x- and y-axes 
in the pseudo-spin space and transform the Hamiltonian to the conventional form of a 2D Dirac Hamiltonian: $H \to v {\bm \sigma}\cdot {\bm q}$.
However, in what follows we will use (\ref{HP1}), as it is slightly more convenient for the study of the edge states.

The effective SOC arises due to  the heavy-light hole mixing in the wave function (\ref{Se}).
Certainly there are other SOC mechanisms such as Rashba, Dresselhaus, and even direct SOC with the modulating potential.
However, all these mechanisms are relatively weak while the ``ultrarelativistic'' (see above)
 heavy-light hole mixing  can give SOC comparable with the kinetic energy.
The heavy-light hole mixing in the wave function (\ref{Se}) leads to the following SOC correction to the matrix element of the potential (\ref{uh}):
\begin{eqnarray}
\label{exA}
\delta(\langle {\bm p}_2|U|{\bm p}_1\rangle) &=&-4iW\alpha^2({\bm p}_1\cdot{\bm p}_2)
([{\bm p}_1\times{\bm p}_2]\cdot s)
\end{eqnarray}
Here ${\bm p}_1-{\bm p}_2 =\pm {\bm G}_i$, and ${\bm s}$ is the effective spin 1/2 introduced in (\ref{Se}). At long wavelengths the leading order contribution for the SOC is given by:
\begin{eqnarray}
\label{ex}
\delta(\langle 2|U|1\rangle) &=&-4iW\alpha^2({\bm K}_1\cdot{\bm K}_2)
([{\bm K}_1\times{\bm K}_2]\cdot s)\nonumber\\
&=&i\frac{2}{\sqrt{3}}\eta s_z \ , \ \ \ \  \eta=\frac{3}{2}\alpha^2K^4 W  > 0 \ ,
\end{eqnarray}
which can be written as:
\begin{eqnarray}
\delta H_{ls} &=&
 \frac{2}{\sqrt{3}} \eta s_z \left(
\begin{array}{ccc}
0 & i & -i\\
-i & 0 & i\\
i & -i & 0
\end{array}\right) \, .
\nonumber
\end{eqnarray}
Projecting this matrix to the states $|a\rangle$ and $|b\rangle$ defined by (\ref{cc}) we get: 
$\delta H_{ls} \to -2 \eta  s_z \sigma_y$. Thus, the final Hamiltonian, including (\ref{HP1}), reads:
\begin{eqnarray}
\label{HP5}
H=v\left(-\sigma_zq_x-\sigma_xq_y\right) - 2 \eta  s_z \sigma_y \ .
\end{eqnarray}
The Hamiltonian is written for the {\bf K} Dirac cone.
Under the parity reflection, ${\bm K} \to {\bm K}'=-{\bm K}$,
the kinetic energy (\ref{mev}) changes its sign while the SOC (\ref{ex})
is unchanged.
Hence, at the $K'$ Dirac point the effective Hamiltonian differs from
(\ref{HP5}) only by the replacement $v \to -v$.
After a unitary transformation the Hamiltonian (\ref{HP5})
can be written in its conventional form \cite{Kane05}: 
$H \to v {\bm \sigma}\cdot {\bm q} -2 \eta s_z \sigma_z$ leading to a SOC gap given by:
\begin{equation}
\label{dso}
\Delta_{so}=2 \eta  = 3 \alpha^2 K^4 W \approx \frac{16}{9} \left(\frac{d}{L}\right)^4 W \ ,
\end{equation}
which shows that the SOC gap is a very strong function of the $d/L$ ratio. 

\begin{figure}[ht]
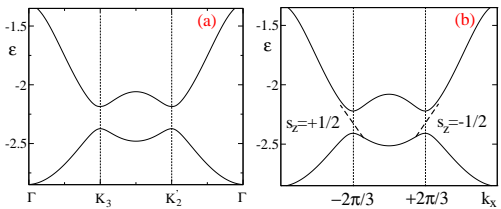

\includegraphics[width=0.18\textwidth,clip]{disp2G.eps}
\includegraphics[width=0.18\textwidth,clip]{EdgeDisp.eps}
\caption{(a) The hole dispersion at ${\overline W}=2$ along the ${\bm \Gamma} \to {\bm K}_2 \to {\bm K}_3^{\prime}
\to  {\bm \Gamma}$ contour in the BZ. The spin orbit splitting corresponds to $d/L=0.5$.
(b) The dashed lines show the dispersion of $s_z=\pm\frac{1}{2}$ edge states. The
solid lines show boundaries of the 2D continuum.
}
\label{dispG}
\end{figure}

SOC  matrix elements (\ref{exA}) can be easily included in the exact diagonalization.
Notice that eqs. (\ref{Se}) and (\ref{exA}) are derived assuming $\alpha^2 p^4 \ll 1$.
Even at small $\alpha$ this condition is violated for high momenta states
included in the exact diagonalization. To avoid this problem we account (\ref{exA}) only
for three lowest quantum states and set the SOC matrix element equal to zero for all higher 
states. This procedure gives the correct gap near the Dirac points.
Notice that, 
$
\alpha^2 K^4  =\frac{16}{27}\ \frac{d^4}{L^4} \approx 0.037 \ ,
$
for $d/L=0.5$. 
The dispersion calculated numerically for this value of the SOC
and ${\overline W}=2$ is shown in Fig.~\ref{dispG}a.
The calculated spin orbit gap is close to the analytical formula (\ref{dso}).
By varying the transverse confinement width $d$, see Eq.(\ref{dso}), one can continuously
change the SOC gap and, hence, the electronic properties of devices made in these systems.

In order to study the presence of edge states we have to consider a sample with a confining potential at the edge.
Having in mind simplicity, we consider an infinite edge potential.
The mechanism of the edge state formation discussed here is qualitatively different from that 
in the graphene \cite{Greview} because of the nature and strength of the potentials in the two cases: graphene is better 
described by the tight-binding model, while semiconductors are better described by the nearly free electron approximation. 
In graphene described by 
the tight binding approximation edge states exist even without the SOC (say, along the zig-zag direction), and the 
SOC only modifies the dispersion of the state~\cite{Kane05}. On the other hand, in the nearly free electron approximation 
the SOC is crucial for the formation of the edge state.

Let us consider the effect of a laterally confining potential to the lattice potential (\ref{uh}). We assume:
\begin{eqnarray}
\label{uconf}
U_{\rm conf}=
\left\{
\begin{array}{ccc}
0 & {\rm if} & y > y_0\\
\infty & {\rm if} & y < y_0\\
\end{array}
\nonumber
\right. \, .
\end{eqnarray}
The envelope wave function of the edge state at $y > y_0$ is given by:
\begin{eqnarray}
\label{bpsi}
\psi=A\left(
\begin{array}{c}
\psi_a\\
\psi_b
\end{array}\right)e^{iq_xx}e^{-\lambda y} \ .\nonumber
\end{eqnarray}
Solving $H\psi=\epsilon\psi$ with Hamiltonian (\ref{HP5}) one finds:
\begin{eqnarray}
\label{0c}
&&\epsilon=\pm\sqrt{\eta^2+v^2q_x^2-v^2\lambda^2}\nonumber \\
&&\psi_a=1 \nonumber\\
&&\psi_b=i\frac{v\lambda+2s_z\eta}{vq_x-\epsilon} \ .
\end{eqnarray}
The corresponding full coordinate wave function reads:
\begin{eqnarray}
\label{psif}
&&\psi\propto e^{iq_xx}e^{-\lambda y}\left\{|a\rangle+\psi_b|b\rangle\right\}\nonumber\\
&&|a\rangle = \frac{1}{\sqrt{2}}\left(e^{i{\bm K_2}\cdot{\bm r}}-e^{i{\bm K_3}\cdot{\bm r}}\right)\nonumber\\
&&|b\rangle = \frac{1}{\sqrt{6}}
\left(2 e^{i{\bm K_1}\cdot{\bm r}}-e^{i{\bm K_2}\cdot{\bm r}}-e^{i{\bm K_3}\cdot{\bm r}}\right) \ ,
\end{eqnarray}
where we have used  Eqs. (\ref{psi}), (\ref{cc}) for $|a\rangle$ and $|b\rangle$.
The wave function must be zero at $y=y_0$ at the position of the confining wall. 
It is obvious that one cannot satisfy this condition at arbitrary $y_0$. 
Fortunately, it is very easy to find the state if the wall position is chosen as: 
$\frac{2\pi}{\sqrt{3}L}y_0=\pi N$ \ , where $N$ is integer.
At $y=y_0$ the basis wave function $|a\rangle$ is zero at any $x$.
Therefore, to have $\psi(x,y_0)=0$ we need only to impose $\psi_b=0$.
Hence, using (\ref{0c}) we conclude that the edge state exists only at 
$s_z=-1/2$:
\begin{eqnarray}
\label{solK}
 s_z=-1/2 \ ,\ \ \ \lambda =\eta/v \ ,\ \ \  \epsilon=-vq_x \ ,
\nonumber
\end{eqnarray}
which is valid near the K Dirac point.
We already pointed out that  at the $K'$ Dirac point the effective Hamiltonian differs from
(\ref{HP5}) only by the replacement $v \to -v$. The 
edge solution (\ref{0c}) is transformed accordingly.
Therefore, at K' the edge state exists only at $s_z=+1/2$:
\begin{eqnarray}
\label{solKp}
 s_z=+1/2 \ ,  \ \ \ \lambda =\eta/v \ ,\ \ \  \epsilon=vq_x \ . \nonumber
\end{eqnarray}
The dispersion of the edge states with $s_z=\pm\frac{1}{2}$ is shown in Fig.~\ref{dispG}(b).
We found the edge states at a special position of the confining wall.
An explicit calculation at a different wall position/shape is more involved since the calculation
must include admixture of high momentum states to the wave function (\ref{psif}).
However, a 
variation of the wall position/shape does not influence 
the edge states since they are topologically protected.

The edge states support the spin current at the edge of system in the regime of a TI.
In Fig.~\ref{dispG} the energy is given in units of the bandwidth $E_0$, Eq. (\ref{e0}), which
depends on the period of the modulating potential. Present  lithographic techniques
can give the period $L=20-30$ nm in GaAs and $L \approx 10$ nm in Si.
According to Fig.~\ref{dispG} this results in the spin-orbital gap $\Delta_{so}\sim 10$ meV.
By increasing the ratio $d/L$ the gap can be further boosted up by a factor $\sim 2$.
All in all the existing technology can produce  a TI operating at temperature  $10- 100K$.
Improvement of lithographic techniques down to scale $L\approx 5$ nm will open way to
tunable room temperature TI.

A disorder created by charged impurities, if strong, can destroy the miniband
structure. There are two issues related to the disorder, (i) the hole  mean free path, 
(ii) local fluctuations of the Fermi energy.
In  clean GaAs the electron mean free path is about 30$\mu$m.
For holes the mean free path is shorter, but still it is about 
5$\mu$m~\cite{Chen12}, so on this side we are safe,
the superlattice can be larger than 100$\times$100 sites.
To estimate the inhomogeneity of Fermi energy we refer to Shubnikov de-Haas oscillations.
For holes in clean GaAs the oscillations are observed down to magnetic field $0.1$ Tesla~\cite{Chen12}.
This corresponds to the cyclotron frequency $0.05$meV and this is the 
upper limit for the Fermi energy inhomogeneity. So, we are safe here too.

In conclusion, we have shown that it is possible to create robust TI states in ordinary semiconductors with 
strong SOC by quantum confinement and superimposed potentials with hexagonal symmetry. We have shown that 
the SOC gaps can be as large as the heterostructure
bandwidth and can be controlled by varying the confinement potential,
the strength and scale of the superimposed potentials. These systems present amazing flexibility and can
be tuned between completely different regimes such as the Dirac metal, TIs, and standard band insulators. 
Thus, they present an opportunity to study exotic physics in the framework of materials that have been 
the basis of the current semiconductor technology. 

We thank A. R. Hamilton, T. Li, A. I. Milstein, and  O. Klochan for discussions.
OPS acknowledges ARC grant DP120101859.
AHCN acknowledges DOE grant DE-FG02-08ER46512, ONR grant MURI N00014-09-1-1063, and the NRF-CRP award "Novel 2D materials with tailored properties: beyond graphene" (R-144-000-295-281).

\end{document}